\documentclass[12pt]{article}
\usepackage{epsfig}

\def\beq{\begin{equation}}
\def\eeq#1{\label{#1}\end{equation}}
\def\eeqn{\end{equation}}

\def\beqa{\begin{eqnarray}}
\def\eeqa#1{\label{#1}\end{eqnarray}}
\def\eeqan{\end{eqnarray}}

\let\bar=\overbar

\def\Dslash{\not{\hbox{\kern-4pt $D$}}}
\def\dslash{\not{\hbox{\kern-2pt $\del$}}}

\def\msb{{\bar{\ssstyle M \kern -1pt S}}}

\usepackage{fancyhdr,graphicx}
\def\Title#1{\begin{center} {\Large {\bf #1} } \end{center}}

\begin{document}

\Title{Deconfinement transition in protoneutron star cores: Analysis within the
MIT Bag model}

\bigskip\bigskip

%+\addcontentsline{toc}{chapter}{{\it L. Skywalker}}
%+\label{SkywalkerLukeStart}

\begin{raggedright}

{\it Taiza A. S. do Carmo and Germ\'an Lugones \index{Carmo, T.A.S.; Lugones, G.}\\
Universidade Federal do ABC\\
Santo Andr\'e, SP\\
Brazil\\
{\tt Email: taiza.carmo@ufabc.edu.br}}
\bigskip\bigskip
\end{raggedright}

\section{Introduction}

The formation of quark matter during protoneutron star (PNS) evolution is still an open issue in the understanding of compact star physics. The standard scenario for the birth of neutron stars indicates that these objects are formed as consequence of the gravitational collapse and supernova explosion of a massive star. Initially, PNSs are very hot and lepton rich objects, where neutrinos are temporarily trapped. During the first tens of seconds of evolution the PNS evolves to form a cold (T $<$ 10$^{10}$ K) catalyzed neutron star. As neutrinos are radiated, the lepton - per - baryon content of matter goes down and the neutrino chemical potential tends to essentially zero in 50 seconds. Deleptonization is fundamental for quark matter formation inside neutron stars, since it has been shown that the presence of trapped neutrinos in hadronic matter strongly disfavors the deconfinement transition. In fact, neutrino trapping makes the density for the deconfinement transition to be higher than in the case of neutrino-free hadronic matter. As a consequence, the transition could be delayed several seconds after the bounce of the stellar core. When color superconductivity is included together with flavor conservation, the most likely configuration of the just deconfined phase is two-flavor superconducting (2SC) provided the pairing gap is large enough. The relevance of this 2SC intermediate phase (a kind of activation barrier) has been analyzed for deleptonized neutron stars but not for hot and lepton-rich objects like PNSs. In the present paper we shall analyze the deconfinement transition in protoneutron star conditions employing the Massachusetts Institute of Technology (MIT) Bag model in the description of quark matter. For simplicity, the analysis will be made in bulk, i.e. without taking into account the energy cost due to finite size effects in creating a drop of deconfined quark matter in the hadronic environment.

\section{The hadronic phase}

For the hadronic phase we shall use a model based on a relativistic Lagrangian of hadrons interacting via the exchange of $\sigma$, $\rho$, and $\omega$ mesons: a non-linear Walecka model (NLWM) \cite{Walecka,Glendenning1991,Boguta1977} which includes the whole baryon octet, electrons and electron neutrinos in equilibrium under weak interactions. The Lagrangian of the model is given by:
\begin{eqnarray}
{\cal L}  = {\cal L}_{B} + {\cal L}_{M} + {\cal L}_{L},
\end{eqnarray}
\begin{eqnarray}
%%%%%%% line 1
{\cal L} & = & \sum_{B} \bar \psi_B  \bigg[\gamma^\mu  \bigg(i\partial_\mu -   x_{\omega B} \ g_{\omega}  \omega_\mu - x_{\rho B} \ g_{\rho} \ \vec \tau \cdot \vec \rho_\mu \bigg) - (m_B - x_{\sigma B} \ g_{\sigma} \ \sigma)\bigg] \psi_B   \nonumber\\
%%%%%%%  line 2
& + & \frac{1}{2} (\partial_{\mu} \sigma \partial^{\mu} \sigma -m_\sigma^2  \sigma^2) - \frac{b}{3}  m_N (g_\sigma\sigma)^3 -\frac{c}{4}  (g_\sigma \sigma)^4    \nonumber \\
%%%%%%%  line 3
& - &   \frac{1}{4} \omega_{\mu\nu} \omega^{\mu\nu} +\frac{1}{2} m_\omega^2 \omega_{\mu}\ \omega^{\mu} -\frac{1}{4} \vec \rho_{\mu\nu} \cdot \vec \rho\ \! ^{\mu\nu} \nonumber\\
%%%%%%%  line 4
& + & \frac{1}{2}\ m_\rho^2\  \vec \rho_\mu \cdot \vec \rho\ \! ^\mu  + \sum_{L} \bar{\psi}_{L}    [ i \gamma_{\mu}  \partial^{\mu}  - m_{L} ]\psi_{L},
\label{octetlag}
\end{eqnarray}
where the indices $B$, $M$ and $L$ refer to baryons, mesons and leptons respectively, with $B$ = $n$, $p$, $\lambda$, $\Sigma^{+}$, $\Sigma^{0}$, $\Sigma^{-}$, $\Xi^{-}$ and $\Xi^{0}$ and $ L = e^{-}, \nu_e $. The coupling constants are g$_{\sigma B}$ = x$_{\sigma B}$ g$_{\sigma B}$, g$_{\omega B}$ = x$_{\omega B}$ g$\omega_B$ and g$_{\rho B}$ = x$_{\rho B}$ g$_{\rho B}$. The ratios x$_{\sigma B}$, x$_{\omega B}$ and x$_{\rho B}$ are equal to 1 for the nucleons and acquire different values for the other baryons depending on the parametrization. In this paper, we use the parametrization labeled as GM1nh \footnote{GM1 parametrization -- given by Glendenning-Moszkowski -- but we assume no hyperons in matter, only nucleons, electrons and neutrinos.} with the following values: $({g_{\sigma}}/{m_{\sigma}})^{2} = 11.79$ fm$^2$, $({g_{\omega}}/{m_{\omega}})^{2} = 7.149$ fm$^2$, $({g_{\rho}}/{m_{\rho}})^{2} = 4.411$ fm$^2$, $b = 0.002947$, $c = -0.001070$, and the maximum mass is 2.32 $M_{\odot}$.

% (Table \ref{table1}).
%
%\begin{table}[h!]
%\begin{center}
%\begin{tabular}{c|c|c|c|c|c|c}
%\hline
% Label  & $\left({g_{\sigma}}/{m_{\sigma}}\right)^{2}$  & $\left({g_{\omega}}/{m_{\omega}}\right)^{2}$ & $\left({g_{\rho}}/{m_{\rho}}\right)^{2}$ & b   &  c     &   $M_{max}$   \\
%        & [$fm^2$]                                      &  [$fm^2$]                                                &  [$fm^2$]  &     &  & [$M_{\odot}$] \\
%
%\hline
%\quad   GM1   \quad  & 11.79  & 7.149  & 4.411  & 0.002947 & -0.001070  & 1.78  \\
%\quad   GM1nh \quad  & 11.79  & 7.149  & 4.411  & 0.002947 & -0.001070  & 2.32   \\
%\quad   NL3   \quad  & 15.8   & 10.51  & 5.35   & 0.002052 & -0.002651  & 1.95    \\
%\hline
%\end{tabular}
%\caption{Parameters of the hadronic equation of state. For each parametrization we give the maximum mass $M_{max}$ of a hadronic star.}
%\label{table1}
%\end{center}
%\end{table}

\section{The quark matter phase}

The quark phase is composed by $u$, $d$, and $s$ quarks, electrons, electron neutrinos and the corresponding antiparticles. We describe this phase by means of the MIT Bag model at finite temperature with zero strong coupling constant, zero $u$ and $d$ quark  masses and strange quark mass $m_s$ = 150 MeV. The total thermodynamic potential can be written as:
\begin{equation}
\Omega = \sum_{i=q,L} \Omega_{i} + \Omega_{\Delta}  + B ,\label{omega}
\end{equation}%
where the index $q$ and $L$ refer respectively to quarks and leptons. The contribution of the free quarks
is given by:
\begin{eqnarray}
\Omega_{q} = \sum_{f, c} - \frac{g T}{2 \pi^{2}}\int^{\infty}_{0}p^{2}\ln\left[1 + e^{-\left(\frac{E_{cf}-\mu_{cf}}{T}\right)}\right]dp, \label{omega_q}
\end{eqnarray}
with $E_{cf} = \sqrt{p^{2}+ m^{2}_{cf}}$,
being $f  = u, d, s$   the flavor index and $c = r, g, b$  the color index. The contribution of leptons is given by:
\begin{eqnarray}
\Omega_{L} = - \frac{g T}{2 \pi^{2}}\int^{\infty}_{0}p^{2}\ln\left[1 + e^{-\left(\frac{E -\mu }{T}\right)}\right]dp, \label{omega_l}
\end{eqnarray}
with $E = \sqrt{p^{2}+ m^{2}}$. The degeneracy factor is   $ g = 6$ for quarks, $ g = 2$ for electrons and  $ g = 1$ for neutrinos.
The contribution of antiparticles is obtained from Eqs.~(\ref{omega_q}) and (\ref{omega_l}) but with $\bar{\mu}_{cf} = - \mu_{cf}$ and $\bar{\mu} = - \mu$.

For paired quarks we use the expression:
\begin{eqnarray}
\Omega_{\Delta} = \sum_{f, c} - \frac{g T}{2 \pi^{2}}\int^{\infty}_{0}p^{2}\ln\left[1 + e^{-\frac{\varepsilon_{c,f}}{T}}\right]dp, \label{omega_delta}
\end{eqnarray}
with $\varepsilon_{cf} = \pm \sqrt{(E_{cf}-\mu_{cf})^{2}+ \Delta^{2}}$, which is the energy dispersion relation for single-particle when it acquires an energy gap ($\Delta$).
The expression for $E_{cf}$ is the same as we showed above.
%%
%Notice that we can obtain Eq. (\ref{omega_q}) from Eq. (\ref{omega_delta}) in the limit $\Delta = 0$, since we consider both regions: $E_{cf} > \mu_{cf}$ and $E_{cf} < \mu_{cf}$ \cite{Schmitt_2010}.

According to the Bardeen-Cooper-Schrieffer (BCS) theory the temperature dependence of the gap parameter of Eq.~(\ref{omega_delta}) is given by the following expression:
\begin{eqnarray}
%\Delta(T) = \Delta_{0}\sqrt{1- \left(\frac{T}{T_{c}}\right)^{2},} \label{delta(T)}
\Delta(T) = \Delta_{0} \bigg[1- \left(\frac{T}{T_{c}}\right)^{2}\bigg]^{1/2}, \label{delta(T)}
\end{eqnarray}
where the critical temperature is,
\begin{eqnarray}
T_{c} = 0.57 \Delta_{0}. \label{Tc}
\end{eqnarray}

%The Gibbs free energy per baryon is given by,
%%
%\begin{eqnarray}
%g = \frac{1}{n_{B}}\left[\sum_{fc} \mu_{fc} n_{fc} + \mu_{e} n_{e} + \mu_{\nu_{e}} n_{\nu_{e}}\right]. \label{g_quark}%g_{quark} = \frac{\sum_{fc} \mu_{fc} %n_{fc} + \mu_{e} n_{e} + \mu_{\nu_{e}} n_{\nu_{e}}}{n_{B}}. \label{g_quark}%
%\end{eqnarray}
%%%
%where $n_B$ is the baryon number density.

\section{Deconfinement transition in protoneutron stars}

The flavor composition of hadronic matter in $\beta$-equilibrium is different from that of a $\beta$-stable quark-matter drop. Roughly speaking, the direct formation of a $\beta$-stable quark-drop with $N$ quarks will need the almost simultaneous conversion of  $N/3$ up and down quarks into strange quarks, a process which is strongly suppressed with respect to the formation of a non $\beta$-stable drop by a factor $\sim$ $G^{2N/3}_{Fermi}$. For typical values of the critical-size $\beta$-stable drop ($N$ $\sim$ 100 - 1000) the suppression factor is actually tiny. Thus, quark flavor must be conserved during the deconfinement transition.

In order to determine the transition conditions, we apply the Gibbs criteria, i.e. we assume that  deconfinement will occur when the pressure and Gibbs energy per baryon are the same for both hadronic matter and quark matter at a given common temperature. Thus, we have,
\begin{eqnarray}
T^{H} &=& T^{Q} \;\;\; (\textrm{Thermal Equilibrium}), \\
P^{H} &=& P^{Q} \label{dec01} \;\;\; (\textrm{Pressure Equilibrium}),\\
g^{H} &=& g^{Q} \label{dec02} \;\;\; (\textrm{Chemical Equilibrium}), \\
Y_{i}^{H} &=& Y_{i}^{Q}  \label{dec03} \;\;\; (\textrm{Flavor Conservation}),
\end{eqnarray}
%%
%\begin{eqnarray}
%T^{H} &=& T^{Q} \;\;\; (\textrm{Thermal Equilibrium}), \\
%P^{H}(T^{H}, \mu_{p}, \mu_{\nu_{e}}^{H}) &=& P^{Q}(T^{Q}, \{ \mu_{fc} \},  \mu_{e}^{Q}, \mu_{\nu_{e}}^{Q}) \label{dec01} \;\;\; (\textrm{Pressure Equilibrium}),\\
%g^{H}(T^{H}, \mu_{p}, \mu_{\nu_{e}}^{H})&=& g^{Q}(T^{Q}, \{ \mu_{fc} \}, \mu_{e}^{Q}, \mu_{\nu_{e}}^{Q}) \label{dec02} \;\;\; (\textrm{Chemical Equilibrium}), %\\
%Y_{i}^{H}(T^{H}, \mu_{p}, \mu_{\nu_{e}}^{H}) &=& Y_{i}^{Q}(T^{Q}, \{ \mu_{fc} \}, \mu_{e}^{Q}, \mu_{\nu_{e}}^{Q})  \label{dec03} \;\;\; (\textrm{Flavor %Conservation}),
%\end{eqnarray}
%%
being  $Y^H_i \equiv n^H_i / n^H_B$ and  $Y^Q_i \equiv n^Q_i /
n^Q_B$ the abundances of each particle species in the hadronic and quark
phase respectively and $i = u, d, s, e, \nu_e$.

The deconfined phase must be composed by an equal
number or \textit{red}, \textit{green} and \textit{blue} quarks (locally colorless):
\begin{eqnarray}
n_{r} &=& n_{g}, \label{dec04}\\
n_{r} &=& n_{b}. \label{dec05}
\end{eqnarray}
%%
%\begin{eqnarray}
%n_{r}(T^{Q}, \{ \mu_{fr} \}) &=& n_{g}(T^{Q}, \{ \mu_{fg} \}) , \label{dec04}\\
%n_{r}(T^{Q}, \{ \mu_{fr} \}) &=& n_{b}(T^{Q}, \{ \mu_{fb} \}) . \label{dec05}
%\end{eqnarray}
%%

When color superconductivity is included together with flavor
conservation and color neutrality, the most likely configuration of the just deconfined
phase is 2SC provided the pairing gap is large enough \cite{Lugones2005},
\begin{eqnarray}
n_{ur} &=& n_{dg}, \label{dec06} \\
n_{dr} &=& n_{ug}. \label{dec07}
\end{eqnarray}
%%
%\begin{eqnarray}
%n_{ur}(T^{Q}, \mu_{ur}) &=& n_{dg}(T^{Q}, \mu_{dg}) , \label{dec06} \\
%n_{dr}(T^{Q}, \mu_{dr}) &=& n_{ug}(T^{Q}, \mu_{ug}). \label{dec07}
%\end{eqnarray}
%%

In the following we analyze the effect of temperature, neutrino trapping and color superconductivity
on the deconfinement transition for the GM1nh parametrization.

\section{Results}
%
%%Fig.1
%
\begin{figure}[b]%[htb]
\begin{center}
\epsfig{file=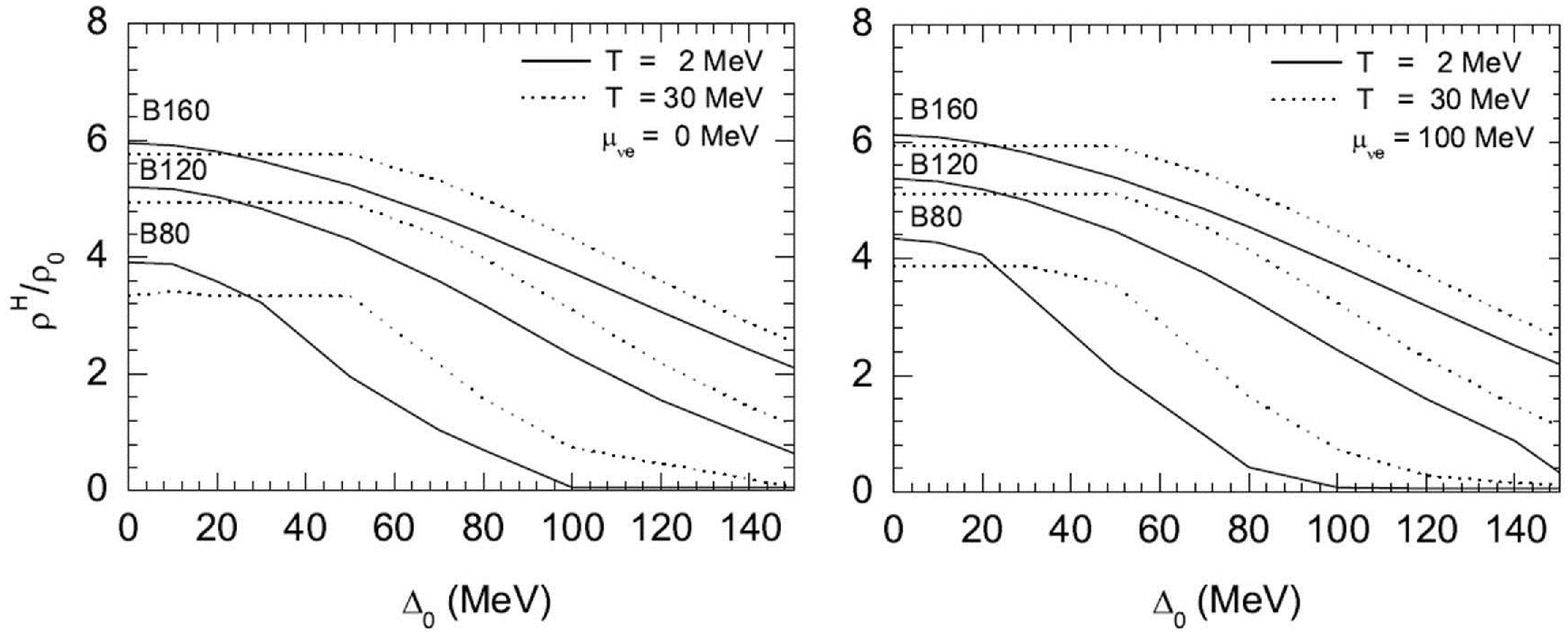,height=2.5in}%scale=0.37
\caption{Mass-energy density of hadronic matter at which deconfinement occurs versus as a function of $\Delta_0$. The values of the bag constant are B $= 80$ MeV/fm$^{3}$ ($B80$), B $= 120$ MeV/fm$^{3}$ ($B120$) and B $= 160$ MeV/fm$^{3}$ ($B160$) \cite{TaizaMIT2013}.}
\label{1}
\end{center}
\end{figure}

%Fig.2  **That figure was not in Event´s presentation.
%
\begin{figure}[b]%[htb]
\begin{center}
\epsfig{file=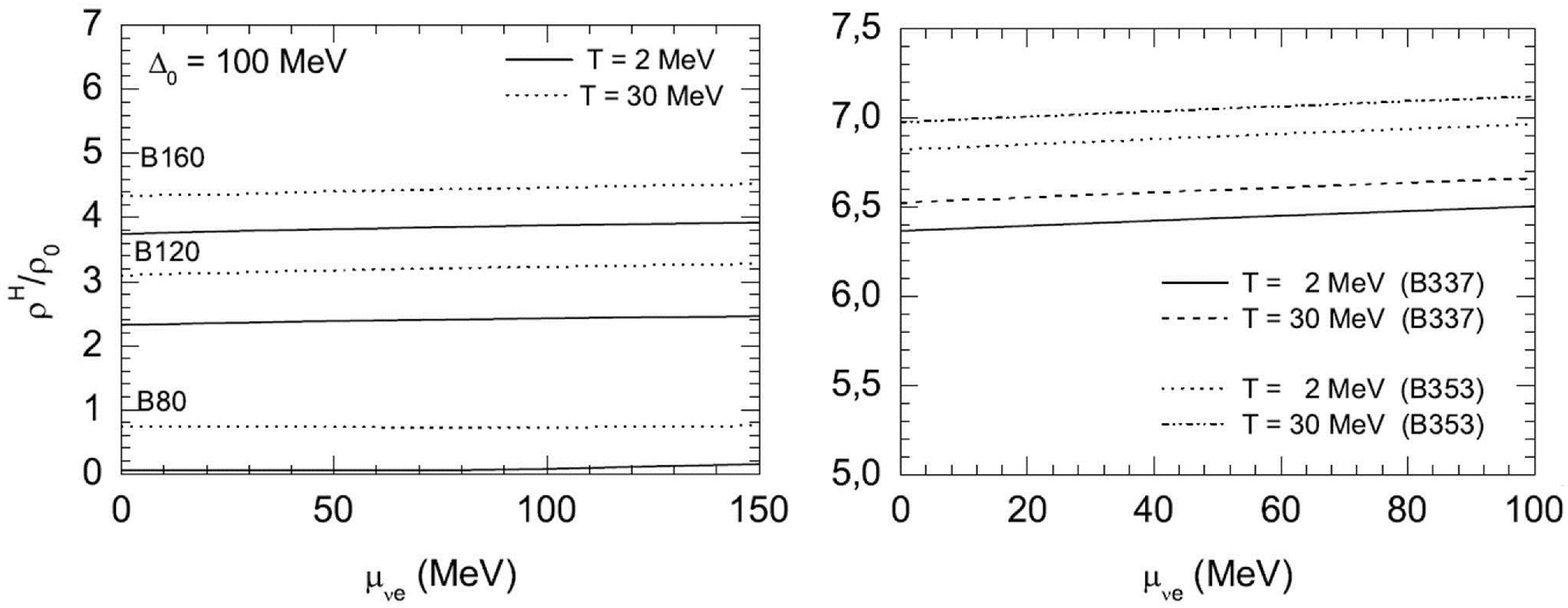,height=2.5in}%
\caption{Mass-energy density of hadronic matter at which deconfinement occurs versus the chemical potential of trapped electron neutrinos $\mu_{\nu e}$ in the hadronic phase \cite{TaizaMIT2013}.}
\label{2}
\end{center}
\end{figure}

In Figure~\ref{1} we show the behavior of the transition density with the gap parameter $\Delta_0$ for the GM1nh parametrization of the hadronic EoS.
The left panel of the figure corresponds to $\mu_{\nu_{e}}^{H} = 0$ MeV and the right panel to $\mu_{\nu_{e}}^{H} = 100$ MeV.
The three full lines correspond to T = 2 MeV and the three dotted curves to T = 30 MeV. Each pair of curves correspond to three different values of the bag constant $B$.  Notice that the  mass-energy density of hadronic matter at which deconfinement occurs is a decreasing function of the gap parameter $\Delta_{0}$.
The effect is strong, e.g. the transition density for $\Delta_0$ = 100 - 150 MeV is much smaller than for $\Delta_0$ = 0 MeV.
For sufficiently small $\Delta_{0}$ the transition density $\rho^{H}$ has constant values. This is because this part of the curve corresponds to temperatures that are larger than the critical temperature $T_{c} = 0.57 \Delta_{0}$, and therefore the pairing gap $\Delta(T)$ is zero.

In Figure~\ref{2} we show the effect of the chemical potential of trapped neutrinos $\mu_{\nu_{e}}^{H}$. The left panel is for $\Delta_{0}= 100$ MeV and the right panel is for different values of bag constant, $B = 353$ MeV/fm$^{3}$  and $B = 337$ MeV/fm$^{3}$. These values of $B$ correspond respectively to the \textit{set 1} and \textit{set 2} parametrizations of the Nambu-Jona-Lasinio (NJL) model employed in \cite{LugonesCarmoNJL}.
Notice that the deconfinement density  is very similar to the obtained in Figure~3 of \cite{LugonesCarmoNJL} within the frame of the NJL model.

In previous work \cite{TaizaMIT2013}, we perform a more detailed study of quark deconfinement in protoneutron stars using color superconductivity in MIT Bag model.

\section{Conclusions}

In this work we analyze the effect of color superconductivity in the transition from hadron matter to quark
matter in the presence of a gas of trapped electron neutrinos. We impose color and flavor conservation during the transition in such a way that just deconfined quark matter is transitorily out of equilibrium with respect to weak interactions. For the hadronic phase we use a parametrization
of a non-linear Walecka model which includes the whole baryon octet. For the quark matter phase
we use MIT Bag model including color superconductivity. In our results we see that the condensation term ($\Delta$) becomes important for temperatures below the critical temperature $T_c$ (see Figure~\ref{1}). This effect may be strong if the superconducting gap is large enough (Figure~\ref{2}).

While neutrino trapping increases the transition density the effect of color superconductivity facilitates the phase
transition. Comparing the present results with those obtained within the NJL model, we find that
the behavior of the transition density ($\rho_H$) as a function of temperature is similar for both models.
In both cases we see that color superconductivity facilitates the transition below $T_c$. With respect
to neutrino trapping, the qualitative behavior of the transition density is opposite in both models.
We find in both cases that the transition density decreases at low temperatures because the pairing
gap increases. According to our results, within the MIT model, when color superconductivity is considered, both cooling and deleptonization of the protoneutron star tends to increase the possibility of deconfinement as the protoneutron star evolves.

\end{document}